\documentclass[12pt]{article}
\usepackage{graphics}

\begin{document}
\pagestyle{plain}

\title{ \bf{ Association between Experiences and Representations: Memory, Dreaming, Dementia and Consciousness } }

\author{ Xiaoqiu Huang \\
Department of Computer Science and Plant Sciences Institute \\
Iowa State University \\
Ames, Iowa 50011, USA }

\date{}
\maketitle

\medskip
\noindent
Running Title: Memory, Dreaming, Dementia and Consciousness

\medskip
\noindent
Corresponding Author:

\medskip
\noindent
E-mail xqhuang@iastate.edu \\
Phone (515) 294-2432 \\
Fax (515) 294-0258


\newpage

\begin{abstract}
The mechanisms underlying major aspects of the human
brain remain a mystery.
It is unknown how verbal episodic memory
is formed and integrated with sensory episodic memory.
There is no consensus on the function and nature of dreaming.
Here we present a theory for governing neural activity
in the human brain. The theory describes the mechanisms
for building memory traces for entities and explains
how verbal memory is integrated with sensory memory.
We infer that a core function of dreaming is to
move charged particles such as calcium ions
from the hippocampus to association areas to primary areas.
We link a high level of calcium ions concentrations
to Alzheimer's disease.
We present a more precise definition of consciousness.
Our results are a step forward in understanding
the function and health of the human brain
and provide the public with ways to keep a healthy brain.
\end{abstract}

\newpage

\section*{Introduction}

The human brain is fascinating but elusive.
The mechanisms underlying major aspects of the brain remain a mystery.
It is unknown how verbal episodic memory
is formed and integrated with sensory episodic memory.
There is no consensus on the function and nature of dreaming.
There is no precise definition of consciousness.
How does behavior affect brain health?
Here we address these questions.

The cognitive neuroscience of memory is
in need of transformation (Nadel et al. 2012).
Memory is attributed to changes in neural synapses by
synaptic plasticity (Martin and Morris 2002).
It is unclear how these changes in different brain areas
are coordinated.
Another question of interest concerns the mechanisms for building
a representation of a facial image of a person in the brain.
Existing episodic memory theories are concerned only with
sensory episodic memory
(Squire 1992; Nadel and Moscovitch 1997; Eichenbaum and Cohen 2001;
Bussey and Saksida 2007; Teyler and Rudy 2007;
Cowell et al. 2010; Nadel et al. 2012); no issues are
addressed concerning the existence and formation of verbal episodic memory.
We describe a theory for governing neural activity
in the human brain, called the theory of associativity.
We address these questions in the framework of the theory.

Memory is linked to sleep as studies show that memory depends
on getting sufficient sleep (Ellenbogen et al. 2006).
Sleep is divided into two types: rapid eye movement (REM) and non-REM.
Most dreaming occurs in REM sleep (Hobson 2009).
A primary function of sleep is thought to consolidate
or transform memory from the hippocampus into the neocortex
(Nadel et al. 2012).
There is no agreement on the purpose and mechanism of
dream construction (Freud 1900;
Hobson and McCarley 1977; Jouvet 1999; Revonsuo 2000).
Existing theories of dreaming could not explain
why dreaming is physiologically and evolutionarily
important or why children have much longer REM sleep than adults.
We develop a theory of dreaming based on the theory of associativity
to address these questions.

Memory is also linked to Alzheimer's disease (AD)
as memory loss is the most common symptom.
AD is characterized by
extreme shrinkage of the cerebral cortex and hippocampus.
AD is linked to disrupted REM sleep (Bliwise 2004).
Quite a few well-known people such as Isaac Newton and Albert Einstein
in the arts and sciences are believed to have the Asperger syndrome (James 2003).
How is intelligence associated with common neural disorders?
We provide new knowledge to explain these links
based on the new theories of associativity and dreaming.

Memory is involved in consciousness as conscious experiences
are encoded and stored by memory mechanisms.
There is no rigorous definition of consciousness, although
theories are proposed to explain consciousness in terms of
neural activities in the brain (Edelman 2004; Koch 2012).
Existing definitions could not
recognize non-human consciousness in other animals or
explain why consciousness evolved as an adaptive advantage.
We present a new definition that is both philosophical and operational.
The definition is used to address these questions.

\section*{ Theory of Associativity }

We expand classical properties of synaptic plasticity
(Bliss and Collingridge 1993; Martin and Morris 2002)
and prevalent views in the cognitive neuroscience of memory
(O'Keefe and Nadel 1978; Squire 1992;
Nadel and Moscovitch 1997; Eichenbaum and Cohen 2001;
Bussey and Saksida 2007; Teyler and Rudy 2007; Cowell et al. 2010)
into a version called the theory of associativity,
which forms the basis for the brain's neural activity.
The theory describes the relationship between
information attended to by the brain
and its neural activity. The theory is described as follows.
There are two major types of information: verbal and non-verbal sensory,
with sensory information further divided into subtypes.
A special example of verbal information is time,
which has no sensory perception.
Neurons are activated by synaptic plasticity to form
synaptic connections when the brain attends to an instance of information
during learning.
A block is a group of connected neurons that are always activated together.
Blocks that are activated at the same time
can be connected by synaptic plasticity into a structure.
How long the structure persists depends on the length and intensity of
attention and emotion as well as the type of neurons involved.
The structure can be strengthened through retrieval or rehearsal.
In general, for some integer $m \geq 0$,
structures at level at most $m$ that are activated at the same time
can be connected into a structure at level $m+1$,
where a structure at level 0 is a block.

The hippocampus is the control center for combining
components into a structure in the neocortex.
The neocortex consists of primary sensory
and sensory association areas, spatial areas,
primary verbal and verbal association areas,
and primary motor and motor association areas.
Note that verbal and visual information processing is
known to involve nonoverlapping brain areas (Newman et al. 2007).
The visual association areas are divided into
the ventral visual stream (VVS) for entity representation and
the dorsal visual stream (DVS) for representation of entity location and motion
(Ungerleider and Mishkin 1982).
The spatial areas are activated to perceive a sketch
of a spatial arrangement of entities in a scene during recollection.
Sensory structures are distributed in sensory association areas;
verbal structures are distributed in verbal association areas.
Association areas are capable of forming structures
by synaptic plasticity in different locations.
However, the hippocampus coordinates the encoding activity in all
the association areas to avoid forming duplicate structures
in different locations for the same instance of information.
Components (in association areas) that are activated at the same time
are first combined into a structure in the hippocampus.
Then the structure is formed in the association areas
by transforming the hippocampal connections between
the components into the association areas.
Sensory hippocampal connections can be transformed through rehearsal,
while verbal hippocampal connections can be immediately
transformed and later strengthened through recollection.
Note that the hippocampus is known to have increased activation
for spatial compared to verbal information (Ryan et al. 2010).

Complex sensory and verbal memory traces for an entity
(individual, class, or background)
are constructed through combination and recursion.
The memory trace for a visual image of a person's face
is constructed in VVS through rehearsal.
The memory trace is a structure at level $n$ for some integer
$n \geq 1$ in VVS.
Assume that the components of this structure, structures at
level at most $n-1$, are already in VVS (Bussey and Saksida 2007).
The assumption is certainly true at $n = 1$,
where the structures at level 0 are blocks in VVS.
(It can be shown similarly that blocks in VVS
can be constructed through combination and recursion
from low-level components in VVS.)
When the visual image is attended to, the components
are activated and combined in the hippocampus according
to the spatial relationship in DVS between the components.
Some of the components are common, whereas the others
are unique ones encoding the specific visual features of the image.
The connections in the hippocampus to the components
are transformed into VVS when the visual image is
repeatedly attended to.
Similarly, the verbal memory trace for a person's name
can be constructed in a verbal association area.

There are two necessary conditions for building
sensory and verbal structures for an entity in the neocortex.
First, the sensory perception of the entity is
different from that of any other entity.
Second, there is a unique verbal reference
to identify the entity.
For example, if the entity is a known person,
then both conditions are satisfied.
If the entity is an unknown person,
then the second condition is not satisfied.
If the entity is a tiger, then both conditions
are not normally satisfied.
If the entity is a class of tigers,
then both conditions are satisfied for the class.
The creation of a unique verbal reference for an entity is
the brain's way of indicating that the entity is worth
attending to. So the brain can ignore some aspects of
the world by not creating unique verbal references for them.
This observation suggests that language evolved for
memory by indicating which aspects of the world are
worth attending to and remembering.

The verbal structure for a verbal reference is formed first.
The structure consists of common components linked by
unique connections, with its meaning
defined as the sensory perception of the entity.
Then the sensory structure for the entity
is formed. The sensory structure also consists of
common components linked by unique connections.
In addition, the unique connections
of the sensory and verbal structures are linked.
The linkage allows the sensory structure to be strengthened
through recollection by activating the verbal structure.
The linkage is also crucial in forming sensory
and verbal episodic memory at the same time.
The sharing of common components in forming structures
is an efficient way of building memory traces
for many known people in the neocortex.
The visual structure for each known person has
unique connections to encode the unique visual
feature of the person.

Consider an example of forming and activating
the connected visual and verbal structures for a female friend.
The connection between the structures is formed during initial encounters
with her in which her verbal name and facial image
are encoded together and therefore the structures are connected.
The connection is strengthened on each subsequent encounter in which
seeing her face and activating the visual structure concur with
calling her name and activating the verbal structure in a greeting.
Thus, seeing her face activates the visual and verbal structures
at the same time, and so does thinking of her name.
This observation is supported by previous findings
(Martin and Chao 2001; Goldberg et al. 2006).

Similarly, there are two necessary conditions for building
sensory and verbal structures for an action.
The sensory perception of the action is
different from that of any other action;
there is a unique verbal reference to identify
the action. The sensory and verbal structures
for a known action such as walking or eating
are connected so that performing or observing the action 
activates both structures.

We are in a position to explain how declarative memory is formed.
Declarative memory is divided into verbal memory and sensory memory.
Both sensory memory and verbal memory each have multiple levels of
organization according to the theory of associativity.
The integration of sensory memory with verbal memory
can be understood in terms of the two human memory properties discussed above.

Consider episodic memory formation at the integration level
when a person attends an event in which the person is
engaged in an activity with a few entities at a time in a place.
Assume that the person is familiar with the entities, place and activity
and that the person already has visual and verbal memory traces for them
at the entity level in association areas.
In order for the person to get to the place on time, the verbal memory
traces for the event and its place and time have to be activated before the event.
Upon arriving at the place, seeing the place activates the visual
and verbal memory traces for the place.
When the person is engaged in the activity with the entities,
engaging in the activity activates its verbal memory trace
and recognizing the entities activates their visual
and verbal memory traces.
At the same time, a new memory trace for the scene
is formed in the hippocampus based on the spatial
information in DVS on the arrangement of the entities.
The new memory trace for the scene represents a spatial
arrangement of the entities in the scene, with each entity
represented by a part in the trace.
The activated visual and verbal entity memory traces are connected
to the corresponding entity parts in the hippocampus.
Then the connections in the hippocampus
enable the fragile connections among the activated term-level verbal memory traces
to be formed in the verbal association areas.
The fragile verbal connections in the verbal association areas
can be strengthened by recollection.
The memory trace for the event consists of a verbal structure with
connections to all the activated entity-level verbal memory traces
as components in the verbal association areas, a visual structure
in the hippocampus with connections to all the activated
entity/action-level visual memory traces as components,
and connections between the two structures in the hippocampus.
The entire memory trace is strengthened through recollection, which
is made by activating the verbal memory traces for the event and its place and time
through the perception of their verbal terms.

In general, a conscious experience is associated with
a superstructure distributed across the hippocampus and association areas.
The superstructure is made up of verbal structures
with components and their connections
in verbal association areas and sensory structures
with components in sensory association areas
and their connections in the hippocampus.
The verbal structures are outside the hippocampus
and are connected to the sensory structures in the hippocampus.
An example superstructure is shown in Figure 1.
An index of the superstructure is a subset of its verbal components such that
its activation triggers the activation of the whole superstructure.
The index is used to recollect the experience
associated with the superstructure.
Note that no superstructure is a component of another superstructure.
Superstructures are different if each of them has its own
unique synaptic connections such that no two of them are activated together.
There are infinitely different superstructures to 
be associated with infinitely different possible experiences. 
Note that it is possible for two different superstructures
to share common verbal or sensory components
because the superstructures are formed at different times.
The sharing of components is efficient in the usage of neurons
and is necessary because all components are singular in the brain.
The association between conscious experiences and
their superstructures is contingent on
the singularity of components and superstructures.

\begin{figure}

\centerline{\scalebox{0.90}{\includegraphics{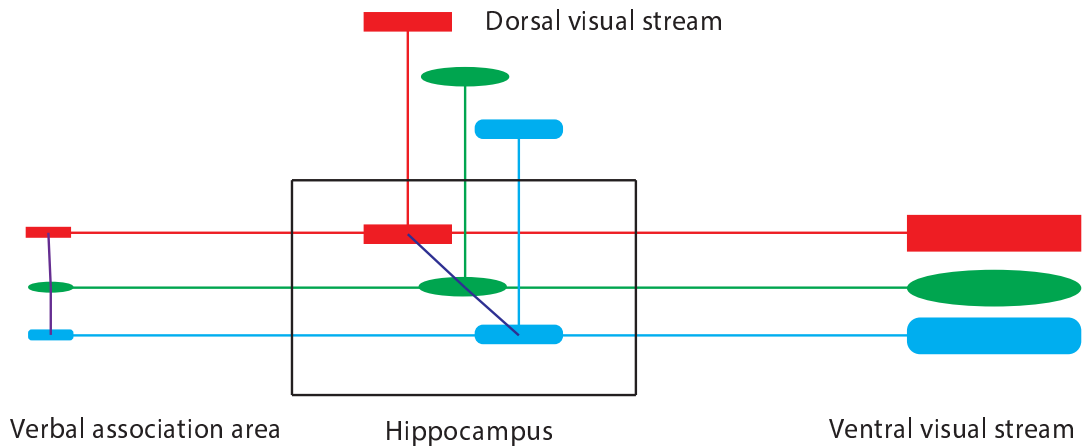}}}

\caption{A memory superstructure formed
when a scene with three entities is attended to.
The superstructure consists of three groups of memory structures
(one group per entity) that are connected together
in both the hippocampus (the open rectangle)
and a verbal association area.
Each group has a verbal structure, a hippocampal structure
(a placeholder), and a visual structure,
which are represented by
small-, medium- and large-sized objects of the same shape,
respectively. The structures along with their connections
in each group are shown in the same color.
Spatial information on the arrangement
of the entities in the scene is passed through
the dorsal visual stream and encoded in the hippocampus.}

\end{figure}

The theory of associativity supports the view
that language, a system of sound-meaning associations,
evolved for memory, which is related to the view of
Hauser et al. (2002) that language evolved for internal thought.
For each entity, its verbal representation (sound)
is connected with its sensory representation (meaning)
for simultaneous activation;
for each experience, its verbal representation (sound)
is connected with its sensory representation (meaning)
for simultaneous activation.
These connections allow the brain to understand precisely the
sensory meaning of each verbal sound.
The sound is an efficient and permanent
way to represent the meaning;
the meaning is replayed back as a recorder
to strength the neural representation for the sound.
The sound is also used to recollect the meaning.
In general, the meaning of a sound is specified as
a unique pattern of brain activation.
A new sound can be recursively defined by specifying
as its meaning a combination of sounds with well-formed meanings.

The theory of associativity is consistent with previous studies
in that episodic memory and semantic
memory are different views of the same underlying neural networks
(Rajah and McIntosh 2005; Burianova and Grady 2007).
The episodic view is obtained by selecting all the parts
representing unique and concrete personal experiences,
while the semantic view is obtained by selecting all the parts
representing general facts and abstract concepts.
Both episodic memory and semantic memory involve
verbal and sensory information.

\section*{ Dreaming }

Dreaming is related to memory processing because a dream can
be recollected the next morning.
Further, dreaming is thought to involve memory activation
(Stickgold et al. 2001).
Thus, an approach to understanding dreaming is by studying neural activation.
Activation is an electro-chemical process by which neurons interact
to transmit information (Bliss and Collingridge 1993).
In the direction of transmission, each of the neurons in turn
opens its membrane channels to let in a flood of calcium ions
as charged particles.
A side effect of activation is that extracellular calcium ions
are pulled in the direction of transmission.
During waking, sensory information constantly reaches
sense organs, resulting in frequent instances of activation.
These instances of activation cause calcium ions to flow in
from primary sensory areas to sensory association areas to the hippocampus.
Thus, at the end of the day, the hippocampus and association areas have
a significantly increased level of extracellular calcium ions than
the primary sensory areas.
Note that before sleep, the hippocampus and association areas
also contain intracellular calcium ions as calcium influx never
ceases during waking.

During dreaming, visual imagery is often perceived in primary visual areas;
evidence for the involvement of primary visual areas is that
blind people whose primary visual cortex is never activated
have no visual imagery in their dreams (Hurovitz et al. 1999)
and that increased activity in the visual cortex
is observed during REM sleep (Braun et al. 1997).
The visual imagery is a result of activation in the hippocampus
and visual association areas.
The activation starts with an encoding operation in the hippocampus
and transmits information from the hippocampus
to the visual association areas to the primary visual areas.
Thus, we propose that a function of dreaming is
to flow calcium ions out from the hippocampus
to association areas to primary areas.

Another approach to understanding dreaming is by studying dream contents.
Dreams are thought to be constructed from memory because
dreams share very little similarity to waking events
and are rarely replays of memory (Stickgold et al. 2001).
The theory of associativity holds that the perception
of sensory information is associated with
the activation of its sensory and verbal
memory traces. For example, the perception of a tiger
in a dream is a result of the activation of the
visual and verbal memory traces for the tiger.
The theory can explain how a dream is constructed.
For example, one occasionally dreams of a fearful event that
one has never experienced but has feared of in one's waking life.
Before the dream, one's memory contains the connected sensory and
verbal memory traces for each of the entities and action in the event.
The verbal memory traces for the terms that refer to the entities and action are
connected as one has expressed fear of such an event in verbal thought.
The dream is triggered by an occurrence of one of the entities in
an unrelated event during the previous day.
The memory trace for the dream is constructed
at the integration level during activation by
connecting the sensory memory traces for the entities and action
in the hippocampus to create a scene of the event.
Then the dream is experienced as a result of the activation.

Virtual experiences are constructed during dreaming by
forming unreal memory. In contrast,
real experiences during waking life lead to the formation of real memory.
Because unreal memory is easily distinguished from real memory,
dreaming causes no interference to real memory.
It appears that memory has a temporal structure
such that real memory is formed in currently
active association areas during waking and
unreal memory is formed in currently
dormant association areas during dreaming.
Our examination of dream contents
offers support for the view that a major function of dreaming
is to redistribute calcium ions without disrupting real memory.
Although unreal memory appears to be of no use,
making unreal memory strengthens
the association between sensory perception and
sensory representation activation.
Emotions in dreams are thought to create large
opposite charges to pull calcium ions in the direction of
information transmission.

Each dream is a unique experience, which
is produced by activating a unique superstructure.
In addition, dreams are the window through which we see the power
of superstructures in representing complex experiences.
These observations support that
the association between experiences and their
superstructures is preserved
both in waking life and in dreaming.
The new view on calcium inflow during waking
and calcium outflow during dreaming
is called the calcium theory.
The calcium theory provides no support for
memory consolidation or transformation during sleep
because it would be confusing
to make unreal and real memory during sleep.

The calcium theory predicts that
the amount of sensory encoding in dreams
is positively related to the amount of sensory encoding
in waking life. The reason is that the amount of
calcium flowing out from the hippocampus to
sensory association areas to primary sensory areas
during dreaming is positively related to the amount of
calcium flowing in during waking life.
It is likely that this relationship is maintained
by using waking sources as dream triggers.
I have many times experienced
dreams that were constructed around dream triggers
encountered randomly during the previous days.
This observation further supports the view that
the contents of dreams are random
but their molecular processes are important
to the brain's function.

Our justification for the function of dreaming also applies
to dreams with only verbal thoughts.
The amount of calcium flowing out from verbal
association areas to primary verbal areas during
dreaming is positively related to
the amount of calcium flowing in during waking life.
If a great deal of thinking without any visual imagery
happens during the day, then dreams at night
are full of verbal thoughts.

Dreams occur mostly during REM sleep.
Sleep begins with a short period of REM sleep,
repeats with a long period of non-REM sleep
and a short period of REM sleep,
and ends with short periods of REM sleep.
Electroencephalography (EEG) waves are known to show that
brain activities during REM sleep and waking life
are similar. The calcium theory holds that waking life and dreaming
are involved in complementary instances of activation.
So EEG wave data provide support for the calcium theory.

The calcium theory provides an explanation of why
a young child spends much more time than an adult in REM sleep.
The young brain is much more involved in sensory and verbal learning
than the mature brain.
As a result, the young brain has higher calcium concentrations
in the hippocampus and association areas than the mature brain.
The young brain needs much more REM sleep to
reduce the higher calcium concentrations.
The calcium theory can also explain 
why an animal needs to sleep at night to shut down
sensory systems completely or partially.
During sleep, the animal's brain flows calcium ions out
to primary areas.

The amount of exposure to sunlight during waking
is known to be positively related to the amount of sleep at night.
The calcium theory also gives an explanation to this relationship.
Attending to objects and scenes in bright sunlight
leads to strong calcium signals in primary visual areas
as light signals are translated into calcium signals,
which produces strong calcium inflow to and influx
in visual association areas and the hippocampus.
Thus, more exposure to sunlight during waking
requires more sleep at night to carry
out calcium efflux and outflow.
Note that it is also possible that more calcium ions
in association areas and the hippocampus produce
more neurotransmitters such as serotonin for sleep.

The amount of physical activities during waking is also known to be
positively related to the amount of sleep at night.
Calcium homeostasis can be used to explain this relationship.
The amount of physical activities
is positively related to the amount of calcium ions
entering muscle cells in the body as calcium influx
induces muscle contractions for physical activities.
The amount of calcium ions entering muscle cells
during physical activities is equal to the amount
of calcium ions pumped out of muscle cells during
rest to maintain calcium homeostasis (Berridge et al. 2003).
Thus, more physical activities during waking amount to less time
to pump calcium ions out of muscle cells during waking
and more time to pump calcium ions out of muscle cells
during sleep.
This relationship suggests communication between
muscle cells and neurons during sleep so that the brain
sleeps long enough to complete calcium efflux in muscle cells.

We sum up a core function of sleep as follows.
The function of sleep is to
move calcium ions from the hippocampus to association areas
to primary areas.
The function consists of two alternate steps of
moving intracellular calcium ions out of neurons
and moving extracellular calcium ions away.
The function of sleep can not be performed during waking
because primary areas continuously transmit calcium ions
to association areas to the hippocampus and calcium influx
constantly occurs.
Another function of sleep is to move calcium ions
out of muscle cells in the body so that they can enter
muscle cells during waking.

\section*{ Dementia, Headaches and Relaxation }

Alzheimer's disease (AD) patients have reduced REM sleep in
proportion to the extent of their dementia (Bliwise 2004).
Dreams occur mainly in REM sleep.
Thus, AD patients have fewer and shorter dreams,
and have increased calcium concentrations in the hippocampus
and association areas, according to the calcium theory.
We conclude that widespread neuron death in
the neocortex and hippocampus is linked to
increased calcium ions concentrations in these brain areas.
This conclusion is not surprising as
increased calcium ions concentrations are known to
kill cells (Orrenius et al. 2003).
The new finding linking increased calcium ions concentrations to AD
supports the view that calcium as a neural signaling molecule
has a role in generating common neural disorders
(Krey and Dolmetsch 2007; Marambaud et al. 2009).

The calcium theory also links calcium to common headaches.
After a day of work or study involving calcium inflow
in the brain, the brain feels dizzy and sleepy.
However, after a good night's sleep involving
calcium outflow, the brain feels fresh.
The dizziness and sleepiness is a warning sign from
the brain in response to increased calcium ions concentrations
in the hippocampus and association areas.
Thus, calcium is linked to headaches.
The new finding is consistent with previous findings
linking calcium signaling to migraine (Gargus 2009).

The calcium theory gives an explanation of a personal mystery.
When I constantly pay attention to people speaking in a faculty meeting
for one and a half hours, I get a headache immediately after the meeting.
However, at the coming night, I get a better night's sleep with
a lot of dreams. The headache disappear completely the next morning.
This mystery happens to me every time I attend a meeting.
According to the calcium theory, constant attendance to
people leads to constant activation
involving calcium inflow to and calcium influx in
the hippocampus and visual association areas.
A much higher level of
calcium ions in these areas causes headaches.
At the coming night, sleeping with
the higher level of calcium ions in the hippocampus and
association areas causes the brain to take more sleep time
to get calcium ions out of and away from the neurons in these areas.
On the other hand, if the brain were not able to function
well during sleep, then a buildup of calcium ions in
these areas would occur and cause harm to neurons
because calcium ions are used as a signaling molecule.

This story brings up an important issue of reducing instances
of activation to minimize the occurrences of headaches.
Activation is induced by attention (Newman et al. 2007).
So one way of reducing instances of activation is
to shift attention to something else that is not stressful.
Another way is to relax the brain by making extremely slow
body movements without thinking of or looking at anything.
These techniques work well for me to reduce occurrences of headaches.
In general, relaxation exercises such as Qigong and Tai Chi
are shown to have significant benefits such as
decrease in cortisol levels (Jahnke et al. 2010).
The calcium theory predicts that relaxation exercises
are good for the brain: they inhibit activation,
keep the brain in relaxed states, and help with
sleep by inducing calcium influx in muscle cells.

Calcium ions are involved through activation
in memory, learning and thinking; they are
also involved as a signaling molecule in
common neural disorders.
Thus, calcium ions link intelligence to
common neural disorders.
The ability to remember, learn and think depends
on the ease with which instances of activation occur.
On the other hand, too frequent instances of activation
lead to a buildup of calcium ions in association areas
and the hippocampus, which is harmful to neurons.
In general, calcium ions link waking behavior
to brain health as attending to information and focusing on
thought are part of waking behavior.

A type of waking behavior thought to be beneficial
to brain health is meditation (Cahn and Polich 2006).
The meditating brain can be characterized
by using EEG electrical waves.
EEG waves in the cerebral cortex are classified
into four main divisions named beta, alpha, theta and
delta in a decreasing order of wave frequency.
Beta waves are associated with waking alert mental states;
alpha waves with relaxed mental states;
theta and delta waves often with sleep.
A subtype of beta waves is associated with activation of
receptors of glutamate, an excitatory neurotransmitter of the brain.
Excessive activation of glutamate receptors can cause neural dysfunction
and cell death through excessive calcium influx (Marambaud et al. 2009).
In addition, our common experiences with sleep are that
if the brain is kept in excited states for a long time,
then the brain has difficulty performing its normal function
such as getting into sleep and
staying in sleep for a sufficiently long time.
Meditation helps the brain by training
the brain to transit and stay out of the beta wave state.
Keeping the brain in the alpha wave state for a sufficient
amount of time and in the beta wave state
for a limited amount of time is beneficial to brain health.

\section*{ Consciousness }

The new knowledge that dreaming involves
flowing calcium ions out of the hippocampus and association
areas by forming unreal memory and temporarily paralyzing muscles
suggests that whole sleep be considered a state of unconsciousness.
On the other hand, during waking,
if consciousness is not reflected in behavior that
causes the species to survive, then it is lost,
where behavior includes body movements, communication between
members of the species, and any other instance of motor area activation.
Thus, behavior is part of consciousness that matters.
For example, a predator is avoided by running away or hiding;
prey is actively pursued.
According to the theory of associativity,
if a sensation and a behavior often occur at the same time,
then their representations in the brain are connected by
synaptic plasticity. Therefore, associated sensations
and behaviors concur automatically because they
are wired together in the brain.
Thus, experiences are encoded in
the association between sensory areas and motor areas in the brain.

The above analysis suggests that
consciousness be defined as the ability to behave for
one's own welfare through the association between
sensory areas and motor areas in one's brain.
The theory of associativity explains
the neural mechanisms that underlie consciousness.
The association between a stimulus and a behavior indicates
that the brain not only has a representation of the stimulus,
but also is able to reflect it in the behavior.
The level of consciousness is assessed by measuring
the complexity of behaviors or
the extent of the association between stimuli and behaviors.
Unlike sensations, many forms of behavior can be objectively
observed even for animals.

Other mammals are conscious by the new definition,
which agrees with our intuition.
Studies of their memory show that
the animals have representations for experiences
in their brains. The animals clearly demonstrate
the ability to behave for their own welfare.
It is also easy to conclude from the new definition
that consciousness evolved as a survival advantage.
For example, the ability to discover and remember 
better ways to find foods and avoid danger allows
an individual to gain an adaptive advantage.
Humans have the highest level of consciousness
among animals based on the complexity of their behaviors
such as language behaviors. 

Machines do not have natural consciousness, where
natural consciousness refers to the ability that
is developed by nature.
Machines can be constructed by humans to have synthetic
consciousness when humans understand how their brain works.

\section*{ Discussion }

We have presented a few ideas concerning the
function and health of the human brain.
The new definition of consciousness emphasizes that
behavior matters. It fittingly applies to the brain;
behavior matters a great deal to the function and health of the brain.
For example, sleep is known to be necessary for the brain and memory.
The new calcium theory explains why it is so.
The calcium theory is a by-product of the new theory of associativity
for explaining how verbal memory is formed.
These ideas suggest several directions for future
studies of the human brain.

First, these ideas arrive at a perfect time as
the Brain Research through Advancing Innovative Neurotechnologies
(BRAIN) Initiative is in the process of being launched (Insel et al. 2013).
The theory of associativity gives a high-level
description of how the brain forms representations
of the outside world.
One goal of the BRAIN Initiative can be to collect experimental
data to confirm, revise, or expand the theory.
Another goal can be to use neural calcium imaging technology
to study the distribution of intracellular and extracellular
calcium ions in the primary areas and association areas
and hippocampus before and after sleep.
A third goal can be to find out the role of calcium as
a second messenger in generating common neural disorders and headaches.

Second, it is important to find effective ways of
keeping the brain in a highly healthy and functional condition
over a lifetime. Tactile massage and relaxation exercises are known
to reduce stress. Do these activities on a daily basis
relax the brain by reducing calcium ions in the hippocampus
and association areas? On the other hand,
people are now bombarded with information by watching TV,
surfing the Internet, playing games, and using smartphones.
People are also required to be productive at work by constantly
attending and responding to fast-changing events and situations.
Do these daily activities over long hours cause damage to
the brain by building up calcium ions in the hippocampus
and association areas?
This question is specially important for brains that
are prone to activation.

Third, getting a good night's sleep is an effective way of
keeping a healthy brain. Strength exercises in the morning and afternoon
are known to help with night's sleep. However,
idling the brain and inducing no calcium inflow
during the day do not lead to a good night's sleep
because there is no work to do during sleep according to the calcium theory.
Because dreams tend to be constructed around waking sources,
getting sensory and verbal information during the day
facilitates the construction of dreams at night.
As light signals received by the eyes are translated into calcium signals
in the brain, the type and intensity of light exposed during the day
have an impact on night's sleep.
How do the amount and type of information received in waking
along with environments affect night's sleep?
As sleep depends on certain biochemical reactions,
it is necessary to eat a variety of foods during the day
that contain the chemicals required for sleep.
What kinds of foods help with night's sleep?

The human brain is the most valuable natural resource on the earth.
The ideas proposed in this paper suggest that behavior affects the function
and health of the brain through processes involving
calcium as a second messenger.
It is important to take the best care of the brain with 
healthy behaviors over a lifetime because it is more
difficult if ever possible to reverse the damage done
to the brain with unhealthy behaviors.
Although more studies on the effects of behaviors on the brain are needed,
warning signs such as headaches from the brain are valuable indicators
of the effects of unhealthy behaviors on the brain.

\end{document}